\documentclass[sigconf]{acmart}

\AtBeginDocument{%
  \providecommand\BibTeX{{%
    \normalfont B\kern-0.5em{\scshape i\kern-0.25em b}\kern-0.8em\TeX}}}

\usepackage{booktabs}
\usepackage{gensymb}
\usepackage{multirow}
\usepackage{fontawesome5}
\usepackage{enumitem}
\usepackage{ulem}
\usepackage{tabularx}
\usepackage{colortbl}
\usepackage{soul}
\usepackage{tabularray}

\copyrightyear{2024}
\acmYear{2024}
\setcopyright{acmlicensed}\acmConference[C\&C '24]{Creativity and Cognition}{June 23--26, 2024}{Chicago, IL, USA}
\acmBooktitle{Creativity and Cognition (C\&C '24), June 23--26, 2024, Chicago, IL, USA}
\acmDOI{10.1145/3635636.3656189}
\acmISBN{979-8-4007-0485-7/24/06}

\begin{document}

\title{MIMOSA: Human-AI Co-Creation of Computational Spatial Audio Effects on Videos}

\author{Zheng Ning}
\authornote{Both authors contributed equally to this work.}
\affiliation{%
  \institution{University of Notre Dame}
  \city{Notre Dame}
  \state{Indiana}
  \country{USA}}
\email{zning@nd.edu}

\author{Zheng Zhang}
\authornotemark[1]
\affiliation{%
  \institution{University of Notre Dame}
  \city{Notre Dame}
  \state{Indiana}
  \country{USA}}
\email{zzhang37@nd.edu}

\author{Jerrick Ban}
\email{jban@nd.edu}
\affiliation{%
  \institution{University of Notre Dame}
  \city{Notre Dame}
  \state{Indiana}
  \country{USA}
}

\author{Kaiwen Jiang}
\email{k1jiang@ucsd.edu}
\affiliation{%
  \institution{University of California San Diego}
  \city{La Jolla}
  \state{California}
  \country{USA}
}

\author{Ruohong Gan}
\email{ruohongg@andrew.cmu.edu}
\affiliation{%
  \institution{Carnegie Mellon University}
  \city{Pittsburgh}
  \state{Pennsylvania}
  \country{USA}
}

\author{Yapeng Tian}
\email{yapeng.tian@utdallas.edu}
\affiliation{%
  \institution{University of Texas at Dallas}
  \city{Richardson}
  \state{Texas}
  \country{USA}}

\author{Toby Jia-Jun Li}
\email{toby.j.li@nd.edu}
\affiliation{%
  \institution{University of Notre Dame}
  \city{Notre Dame}
  \state{Indiana}
  \country{USA}}

\begin{abstract}
Spatial audio offers more immersive video consumption experiences to viewers; however, creating and editing spatial audio often expensive and requires specialized hardware equipment and skills, posing a high barrier for amateur video creators. We present \textsc{Mimosa}, a human-AI co-creation tool that enables amateur users to computationally generate and manipulate spatial audio effects. For a video with only monaural or stereo audio, \textsc{Mimosa} automatically grounds each sound source to the corresponding sounding object in the visual scene and enables users to further validate and fix errors in the location of the sounding objects. Users can also augment the spatial audio effect by flexibly manipulating the sounding source positions and creatively customizing the audio effect. The design of \textsc{Mimosa} exemplifies a human-AI collaboration approach that, instead of utilizing state-of-art end-to-end ``black-box'' ML models, uses a multistep pipeline that aligns its interpretable intermediate results with the user’s workflow. A lab user study with 15 participants demonstrates \textsc{Mimosa}’s usability, usefulness, expressiveness, and capability in creating immersive spatial audio effects in collaboration with users. 

\end{abstract}



\renewcommand{\shortauthors}{Zheng Ning, et al.}


\begin{teaserfigure}
  \includegraphics[width=\linewidth]{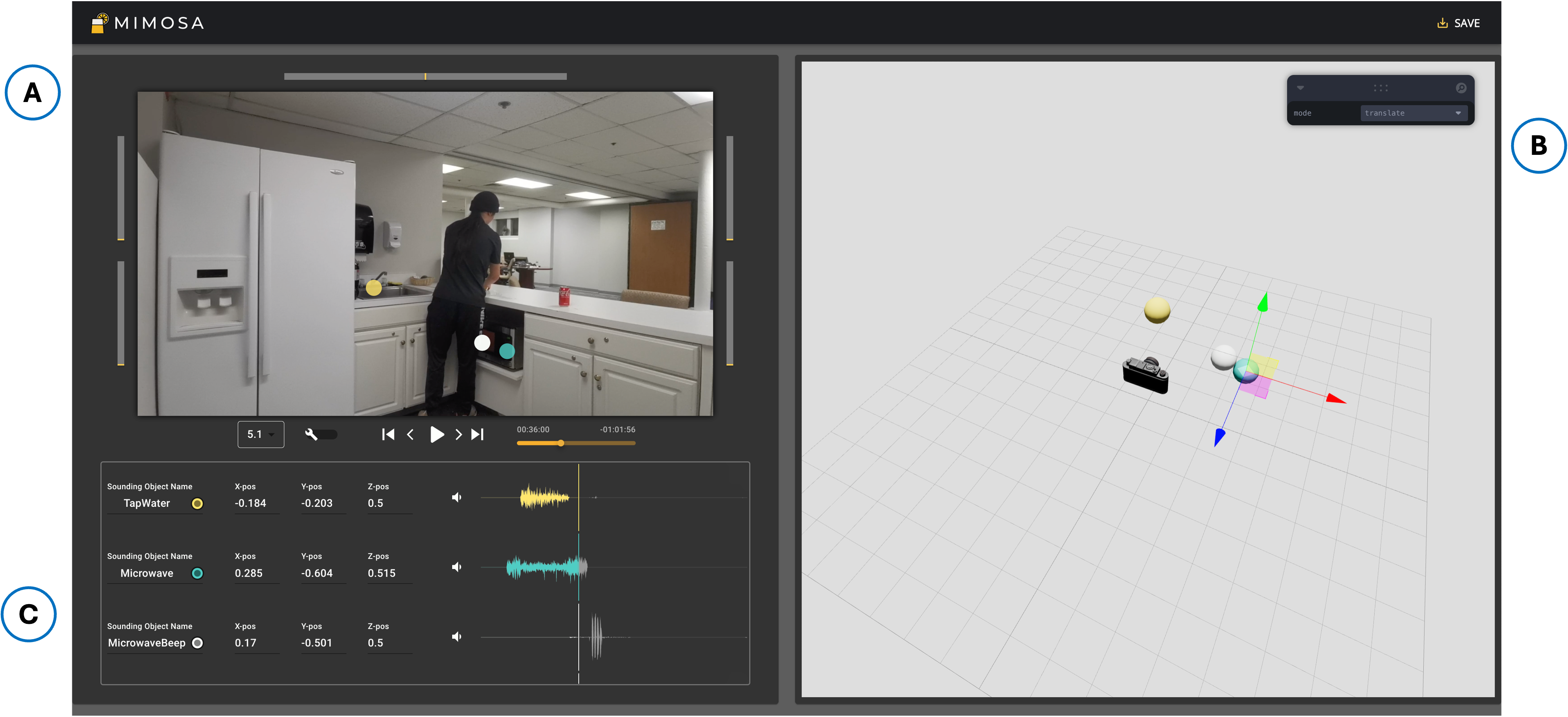}
  \caption{The main interface of \textsc{Mimosa} for spatial audio generation and manipulation}
  \label{fig:mimosa}
  \Description[The main interface of \textsc{Mimosa}]{The main interface of MIMOSA for spatial audio generation and manipulation. The video playback and 2D sounding sources overlay panel are on the upper left of the system UI, the 3D direct manipulation panel is on the right side of the UI, and an information panel that summarizes the meta-information about each individual soundtrack such as sound name, waveform shape, etc. is on the bottom left.}
\end{teaserfigure}

\maketitle

\section{Introduction}
Audio effects play a critical role in enhancing the viewing experience for users. In particular, spatial audio, a type of audio effect that allows listeners to perceive the position of sound sources in a three-dimensional space through surround sound~\cite{rumsey2012spatial, roginska2017immersive}, has been shown to significantly improve the immersion of video content for viewers. Enhanced immersion promotes user engagement, comprehension, and retention of video content~\cite{baldis2001effects, sanchez2003memory, dede2009immersive}.\looseness=-1

However, the adoption of spatial audio among video creators, especially amateurs, remains limited. There are several reasons: first, recording spatial audio requires specialized hardware equipment, such as in-ear microphones (for binaural recording) and 360\degree microphones (for ambisonic recording)~\cite{rumsey2021sound}. Although the equipment has become increasingly affordable, with entry-level ones costing hundreds of dollars, the cost and expertise required for these specialized microphones continue to hinder their widespread adoption. Second, millions of existing videos have been recorded with only monaural or stereo audio~\cite{morgado2018self},  lacking spatial audio information. As a result, content creators cannot directly utilize them as video resources for spatial audio. Lastly, even when the original audio of a video is recorded in a spatial format, post-processing remains challenging and requires specialized tools and expertise~\cite{coleman2017object, mccormack2019sparta}.

Although end-to-end machine learning (ML) models have made significant progress in generating spatial audio from videos with monaural or stereo audio~\cite{morgado2018self,gao20192,zhou2020sep}, they are not yet sufficiently useful for meeting the \textit{practical} needs of content creators due to the following reasons:
\begin{itemize}
    \item [(i)] The creation of audio effects for videos is a personalized process where the quality of the effect is evaluated by the creator's subjective perception~\cite{smaragdis2009user,lin2021soundify}; however, the results generated from the ML models are usually evaluated on ``ground-truth'' quantitative metrics~\cite{morgado2018self,gao20192,zhou2020sep}. The ``black-box'' nature of the state-of-the-art end-to-end ML models prevents users from easily validating and revising their output, as they lack the capability to produce meaningful intermediate results that end users can understand, validate, and edit. \looseness=-1

    \item [(ii)] Existing ML models are primarily designed to reconstruct the ``ground truth'' audio; however, video creators often seek to tailor effects to their personal preferences. For instance, the workflow often includes adjusting volume for each sound source separately and changing the spatial perception by redistributing the sound in a 3D space~\cite{langford2013digital,middletonmix}. They may also experiment with various audio settings~\cite{dalton2016rondo360}. In contrast, ML models are limited to predicting spatial audio that closely approximates how the original audio would sound if it had been recorded with spatial audio equipment.
\end{itemize}

Therefore, using the models alone may not always be the best option in various scenarios. An appropriate tool should support users in continuously exploring the creative space for immersive audio experiences, while providing sufficient flexibility and expressiveness for content creators to realize their ideal audio experiences.

To this end, we introduce \textsc{Mimosa}\footnote{\textsc{Mimosa} is an acronym for \textbf{M}agnifying \textbf{I}mmersion by \textbf{M}anipulating \textbf{O}bjects in \textbf{S}patial \textbf{A}udio} (Fig.~\ref{fig:mimosa}), a human-AI collaborative tool that helps amateur content creators create immersive spatial audio effects for videos with conventional monaural or stereo audio. Rather than relying on an end-to-end ``black-box'' machine learning approach, \textsc{Mimosa} employs a carefully designed audiovisual pipeline (Fig.~\ref{fig:ml_pipeline}) compromising object detection, depth estimation, soundtrack separation, audio tagging, and spatial audio rendering modules to produce useful intermediate results. Those results, such as the type and position of independent soundtracks of different sounding objects, the estimated 3D position of the sounding object that changes with time, are organized and presented to users through an interactive direct manipulation interface, which allows them to easily manipulate the spatial location and audio attributes of each sound source. To further facilitate real-world deployment of \textsc{Mimosa}, we developed an extension for Adobe Premiere Pro, a popular video editing application among our target users. This extension seamlessly integrates \textsc{Mimosa} with Premiere Pro, allowing users to directly invoke \textsc{Mimosa} while editing videos.

We assessed the quality of spatial audio effects generated by \textsc{Mimosa} through a subjective evaluation conducted by eight independent external evaluators. The evaluators reviewed and rated videos featuring various types of audio effects (including the generated spatial audio by using \textsc{Mimosa} and monaural or raw audio as comparisons) without prior knowledge of the type of audio effect being presented. The result demonstrates that videos featuring spatial audio effects produced by \textsc{Mimosa} were more immersive than the original video sound while maintaining a high degree of realism. 

The usability and usefulness of \textsc{Mimosa} is evaluated through a user study with 15 participants who are either amateur content creators or users without prior experience in video or audio editing. Qualitative insights are discussed regarding how users utilize \textsc{Mimosa}'s audiovisual interface to assist their audio editing process.

In summary, this paper makes the following contributions:
\begin{itemize}
     \item \textsc{Mimosa}, an interactive system that enables amateur content creators to create spatial audio effects for videos with monaural or stereo audio only.
     \item A subjective evaluation (N=8) that evaluates the efficacy of the step-by-step spatial sound generation pipeline against the end-to-end models.
    \item Insights from a user study with 15 participants to create spatial audio effects using \textsc{Mimosa}.
\end{itemize}
\section{Related work}
\textsc{Mimosa} builds upon prior research in three key areas. First, we review computational techniques for modeling spatial audio, including approaches for simulating sound propagation and localizing sound sources in videos (Section~\ref{sec:rw_tech}). Next, we discuss the diverse applications of spatial audio, which aim to either accurately position sounds in 3D space or foster an immersive audio experience (Section~\ref{sec:rw_apps}). Finally, we situate \textsc{Mimosa} within the broader context of AI-enabled co-creation tools for multimedia, and highlight key design strategies that informed the development of our tool (Section~\ref{sec:rw_tools}).

\subsection{Computational Techniques for Spatial Audio Generation}
\label{sec:rw_tech}
Spatial audio generation is a broad topic that contains multiple individual research questions. One of them is to model changes in head-related transfer function (HRTF) as sound waves propagate. The HRTF describes how sounds are modified due to bouncing, scattering, and diffraction as they approach and enter the ears of a listener~\cite{gardner1995hrtf,moller1995head}. The modeling of HRTF and sound propagation is based on a large amount of sound source and environmental data. In this scenario, deep learning networks were often adopted to model the environment~\cite{yamamoto2017fully}, human head and ear geometries~\cite{zhang2021personalized,richard2020neural,gebru2021implicit} and the change in sound features~\cite{chaitanya2020directional,raghuvanshi2014parametric}

Another complex challenge arises when dealing with monaural audio recordings from videos that have a mixture of sounding objects. To generate spatial audio effects for such videos, current state-of-the-art approaches often leveraged deep neural networks to associate each independent sound source to their corresponding visual counterparts, and spatialize the video soundtrack with the additional information from the visual objects at different video frames~\cite{morgado2018self, gao20192,zhou2020sep,xu2021visually}.\looseness=-1

In our work, we focus on assisting amateur content creators produce realistic spatial audio effects from videos by allowing users to iteratively validate and edit the intermediate outputs. Furthermore, our work targets at allowing the users to augment the spatial effects for more immersive experience or express their creative goals of the audio effects in the video context. \looseness=-1

\subsection{Applications of Spatial Audio}
\label{sec:rw_apps}
The application of spatial audio can be broadly categorized into two groups based on its aim: (i) accurate 3D positioning and (ii) immersive illusion perception~\cite{rumsey2012spatial}. In the first category, spatial audio provides essential cues that help convey the locations and movements of sound sources. This has been applied in applications such as understanding the positions of multiple speakers~\cite{baldis2001effects,nguyen2017collavr}, augmenting spatial awareness for vision-impaired individuals~\cite{ning2024spica,jain2023front}, aiding in navigation for the blind~\cite{heller2018navigatone}, and enriching object perception in augmented reality environments~\cite{yang2019audio}. The second category focuses on enhancing viewer immersion by creating a spatial audio sound field, which does not necessarily require precise localization of sound sources. Examples include designing the soundtrack of movies to foster an immersive cinema or home theater experience~\cite{clark1957the,north2008the}, and enhancing audio perception in AR/VR environment~\cite{yong2018using}.

Despite spatial audio has demonstrated utility across various application domains, its widespread adoption remains limited due to barriers in creator expertise and equipment availability. A key goal of Mimosa is to mitigate these barriers, promoting more extensive use of spatial audio in diverse application domains.

\subsection{AI-Enabled Co-Creation Tools for Multimedia Experiences}
\label{sec:rw_tools}
\textsc{Mimosa} is part of an expanding group of AI-enabled co-creation tools designed for multimedia experiences. These tools generally aim to achieve two objectives in assisting content creators: (i) improving efficiency, lowering barriers, or reducing effort required in the content creation process; and (ii) expanding the flexibility and expressiveness of the creations~\cite{chung2021intersection,frich2019mapping,remy2020evaluating}. While \textsc{Mimosa}'s primary design goal aligns with the first category, it also contributes to the second category by enabling users to ``go beyond the ground truth'' through manipulation of inferred sounding object positions, ultimately creating customized spatial audio effects.

The design of \textsc{Mimosa} is informed by key design strategies and implications from previous human-AI co-creation efforts in various application scenarios, such as sketching~\cite{lin2020your}, music creation~\cite{simon2008mysong, huang2018music}, and video creation~\cite{lin2021soundify}. Specifically, it ensures that users maintain control and consistently play a ``leading role'' in the co-creation process~\cite{oh2018lead}, fosters accurate mental models of the AI system by employing a pipeline that mirrors the cognitive and reasoning processes of human users~\cite{guzdial2019friend,zhang2023visar,gebreegziabher_patat:_2023, zhang2023peanut}, and allows users to efficiently handle errors~\cite{ning2023empirical,10.1145/319382.319398,10.1145/302979.303163} during the process.

\section{The MIMOSA System}
\label{sec:system}

\begin{figure*}[!htb]
\centering
  \includegraphics[width=0.95\linewidth]{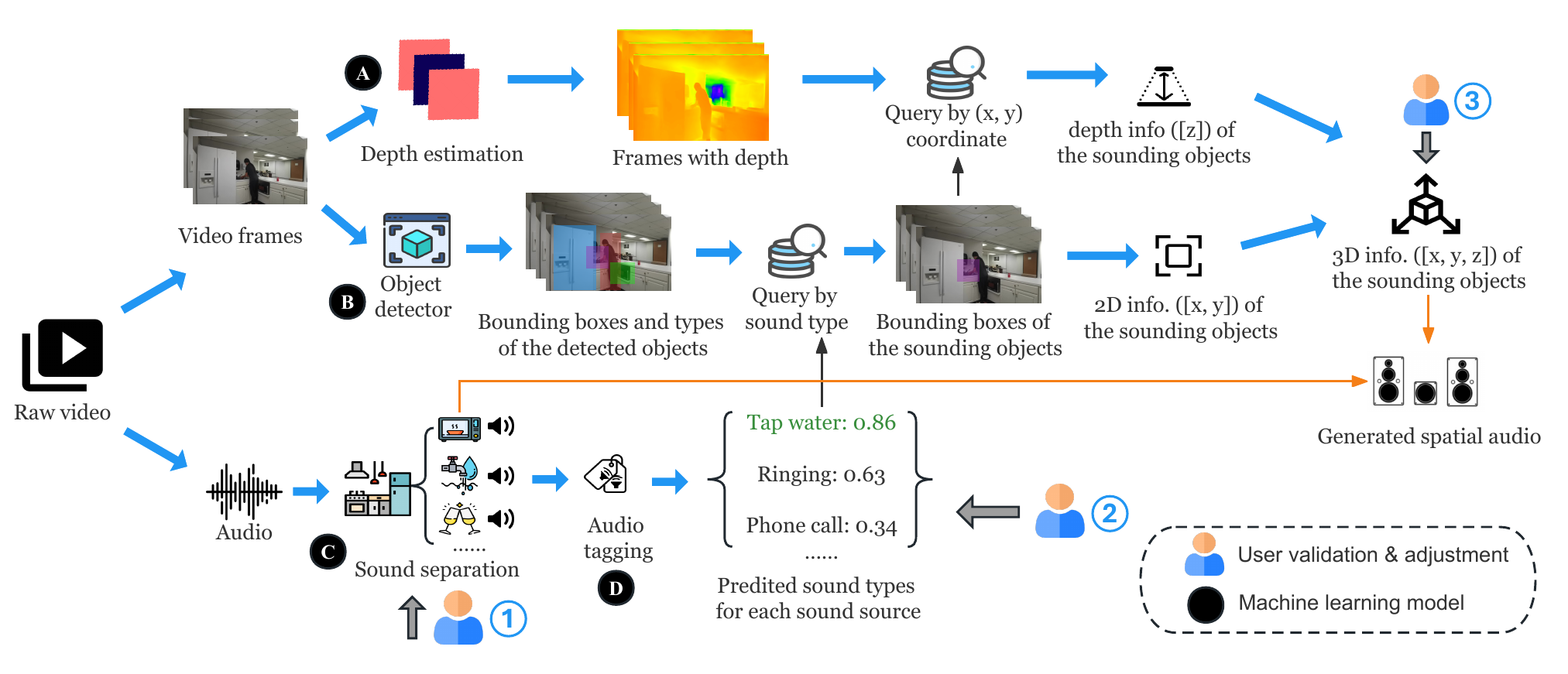}
  \caption{\textsc{Mimosa}'s human-AI collaborative audio spatialization pipeline. Users can validate and adjust the intermediate results in three ways. From left to right, 1: users can adjust the audio properties of each separated soundtrack; 2: users can manually fix the error in aligning the separated soundtrack to the visual object in the video; 3: users can customize the spatial effect for each sounding object by manipulating its corresponding visual position.}
  \label{fig:ml_pipeline}
  \Description[]{The figure shows MIMOSA's human-in-the-loop audiovisual spatialization pipeline. The 3D position of the sounding object is acquired by a set of computer vision models and an audio tagging model telling which object is making the sound. The independent soundtrack of each sounding object is separated from the original soundtrack.}
\end{figure*}

We designed and implemented \textsc{Mimosa}, a human-AI collaborative tool for generating and manipulating spatial audio effects on videos. In this section, we start by outlining the key design challenges and goals of \textsc{Mimosa}. Then, we introduce the architecture of the system (Section~\ref{sec:arch}), an example application scenario (Section~\ref{sec:exp_scene}), the key interactive features (Section~\ref{sec:features}), and the back-end spatial audio generation pipeline (Section~\ref{sec:voasFramework}). This section ends with a discussion of the system's implementation details (Section~\ref{sec:implementation}).

\subsection{Technical Challenges and Design Goals}
\label{sec:challenge}
Informed by previous studies on creativity support tools~\cite{chung2022artist,chung2021intersection,frich2019mapping} and computational approaches in audiovisual context~\cite{wei2022learning}, we identify the following technical challenges (TC) for designing and building \textsc{Mimosa}.

\begin{itemize}

\item[TC1] \textit{Limitations in the performance and capabilities of existing fully-automated ML models.} As discussed in Section~\ref{sec:rw_tech}, current state-of-the-art ML models have limited performance in generating realistic and immersive spatial audio from videos containing only monaural or stereo audio. The quality of the generated spatial audio is significantly affected by the training data, video quality, and video scenarios.

\item[TC2] \textit{Lack of user control in the generation process of spatial audio.} Due to the black-box nature of current end-to-end models, users cannot easily repair the errors in the generated effect, or control the generation process to customize the final output.

\item[TC3] \textit{Increased information overload and cognitive burden for users working with multimodal data}. As users have to handle both audio and visual data while editing the audio effect for a video, the discrepancies divert users' attention between modalities, making it more difficult for them to manage the task at hand~\cite{rubinstein2001executive}. This issue becomes more pronounced when visual and audio information are inconsistent in terms of their respective positions.
\end{itemize}

In light of these technical challenges, we identify three design goals for \textsc{Mimosa}.

\begin{itemize}
\item[DG1] \textit{Allow users to effectively repair the errors in the model-generated spatial audio effects.} To address the limitations of end-to-end models outlined in TC1, the system should provide user-interpretable intermediate results of ML models, allowing users to verify the results generated by the model and rectify errors.

\item[DG2] \textit{Provide flexibility for users to augment the audio effect for a more immersive experience and creatively modify the spatial audio effects.} Instead of strictly follow the visual relative position of the sounding object to spatialize the audio, the system should also allow users to augment the spatial audio effect. Additionally, it should offer users ample expressiveness, such as flexibly moving the individual sound sources and adjusting the reference points to enable them experiencing different audio settings.

\item[DG3] \textit{Coordinate user perceptions, cognition, and reasoning while users are handling audiovisual data.} The system should provide effective and easy-to-learn interaction strategies to help reduce the cognitive load when users edit the audio effects in the video context and using the interactive strategies and UI components of the system.\looseness=-1
\end{itemize}

\subsection{System Architecture}
\label{sec:arch}
The architecture of \textsc{Mimosa} consists of two layers. When a video is loaded, the audiovisual spatializing pipeline (illustrated in Fig.~\ref{fig:ml_pipeline}) runs in the backend to process the video and generate a default spatial audio effect. The pipeline also produces intermediate data, enabling users to identify and address any issues in the generation process.

On the second layer, users can modify the intermediate results generated by the machine learning pipeline in the system's main interface (as shown in Fig.~\ref{fig:mimosa}). It consists of three main components: (i) The video playback and 2D sounding source overlay panel (Fig.~\ref{fig:mimosa}-A). The panel allows users to manipulate the inferred position of each individual sound source in a 2D space that refers to the video frame by moving the corresponding overlay colored dots. (ii) The simulated 3D manipulation panel (Fig.~\ref{fig:mimosa}-B). This panel is designed to facilitate user interaction with sound source positioning within a simulated 3D space, it also allows for modifications to the reference point by manipulating the camera object through movements or rotations. (iii) The audio properties display and control panel (Fig.~\ref{fig:mimosa}-C). This panel allows users to verify and manually specify the corresponding pairings between soundtracks and their associated visual objects within the frame, provides users with a comprehensive overview of the audio properties of each individual soundtrack, and enables users to interact with these properties.

We illustrate the design rationales and detailed descriptions of the interface components in Section~\ref{sec:features}.

\subsection{Example Usage Scenario}
\label{sec:exp_scene}
To demonstrate the use of \textsc{Mimosa}, we present a scenario in which an amateur video creator, Lucy (she/her)\footnote{A pseudonym for demonstrating the usage scenario.}, wanted to enhance the soundtrack of a video featuring two music players playing the flute and violin in a room with immersive spatial audio effects. An example screenshot of a video frame was similar to Fig.~\ref{fig: teaser-A}.

Lucy started by launching Adobe Premiere Pro (Pr) and loading the target video. She selected the desired clip for editing and launched \textsc{Mimosa} from the \texttt{Extension} menu in Premiere Pro. The system automatically started processing the video, when it finished, the main interface of \textsc{Mimosa} (Fig.~\ref{fig:mimosa}) appeared, allowing her to proceed with editing the default generated spatial effects.

Next, she played the video with the default generated spatial audio effects to \textit{detect} potential issues. As the video played, the positions of dots (Fig.\ref{fig: teaser-A}-E), spheres (Fig.\ref{fig:teaser-C}-A), and numeric numbers (Fig.\ref{fig:teaser-B}-C) changed automatically over time, illustrating the detected sounding objects' positions. She could control the play/pause, play progress, and move backward/forward 1 second using the control buttons (Fig.\ref{fig: teaser-A}-D), similar to using a typical video player. The real-time volume indicators (Fig.\ref{fig: teaser-A}-A) visualized the volume from different channels to improve her understanding of the spatial effect at specific times. Additionally, she could check if the dots (Fig.\ref{fig: teaser-A}-E) representing the inferred spatial positions of sounding objects aligned with the actual visual objects in the frame.

During playback, Lucy noticed that the spatial sound of the flute is incorrect and wants to \textit{repair} it. There were three types of errors in general. First, if the spatial positions of the flute and violin were reversed, she could re-specify their correspondence by editing the sounding object name in Fig.\ref{fig:teaser-B}-A. Second, if the volume of either instrument were too loud or too soft, she can adjust the volume of each instrument separately using the volume controller (Fig.\ref{fig:teaser-B}-D). Finally, if the correspondence was correct, but the prediction was inaccurate, she could: (i) dragged the dot representing the flute towards the desired direction in the video display panel; (ii) manipulated the sphere corresponding to each sound source in the 3D panel, which also allowed her to adjust the distance from the viewer to the sounding object; or (iii) manually changed the numeric 3D position of the flute object in the audio properties display and control panel.

As a video creator, Lucy wanted to enhance the spatial effect by not strictly adhering to the estimated visual positions of the sounding objects, but instead following her creative ideas. Specifically, in this case, she hoped to increase the distance between the two sound sources, thereby enabling the audience to more distinctly perceive sounds appearing from two separate directions. To achieve this, she could simply move the red dot further to the left and the yellow dot further to the right.

She also wanted to test how different positions of the two players affect spatial effects and how the audience perceived the scene from various angles. To achieve this, she could create desired spatial effects by manipulating the position of each sounding object using the three previously mentioned approaches. Additionally, she could change the viewing point in the 3D panel by adjusting the camera object's position and angle. The spatial audio effects were re-rendered in real-time.

Once satisfied with spatial audio creation, she could press the \texttt{SAVE} button on the interface. The edited spatial audio track for each of the sounding objects will be loaded to Pr automatically, allowing her to continue editing other aspects of the video with the functionalities from Pr.

\subsection{Key Interface Components}
\label{sec:features}
Specifically, we designed the following UI components to enable amateur content creators to repair the errors introduced at each stage of the ML pipeline, expand the design space for customized spatial audio effects, and promote user creativity.

\subsubsection{2D \& 3D sound source manipulation panels}
\label{sec:2d3dpanels}
\begin{figure}[!htb]
    \centering
    \includegraphics[width=0.9\linewidth]{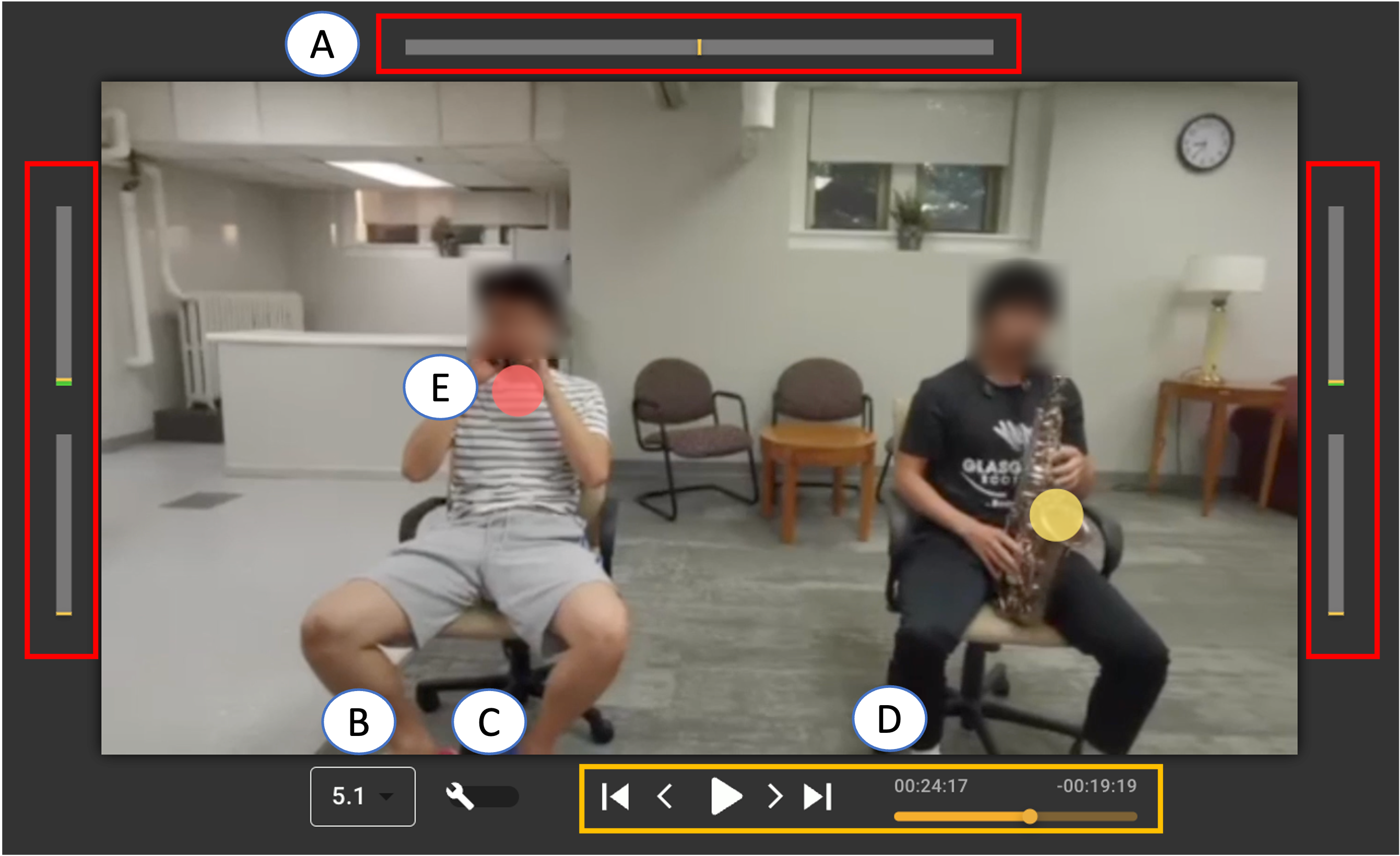}
    \caption{(A) Users can view the volume of each channel using volume indicators. (B) Users can select the audio output format among monaural, stereo, quadraphonic, and 5.1 channels. (C) Users can toggle whether to create spatial effects based on the model-predicted spatial position or their own. (D) Video control buttons, from left to right: previous video, previous second, play/pause, next second, next video. (E) 2D sound source manipulation, where users can adjust the spatial position of each sounding object by moving the corresponding dot.}
    \label{fig: teaser-A}
    \Description[]{The figure illustrates the function of each widget in the system interface. (A) Users can use the volume indicators to see the volume of each channel. (B) Users can change the audio output format among monaural, stereo, quadraphonic, and 5.1 channels. (C) Users can turn on/off the toggle to specify if they want to create the spatial effects based on the model-predicted spatial position or completely on their own. (D) Video control buttons. From left to right: the previous video, the previous one second, play/pause, the next one second, the next video. (E) 2D sound source manipulation. Users can alter the spatial position of each sounding object by moving the corresponding dot}
\end{figure}

The 2D and 3D direct manipulation panels visually represent sound sources as dots in 2D and spheres in 3D within distinct reference systems. These panels enable users to adjust the spatial positions of sound sources to fix errors in the ML model's predictions or create customized spatial effects.

The 2D panel allows users to manipulate sound sources within the original video frame using dot-like visual representations, as illustrated in Fig.~\ref{fig: teaser-A}-E. Each colored dot represents a specific type of sounding object, with its default [x, y] position predicted by the ML model and its size determined by the predicted depth ([z]) from the imaginary camera position. Users can reposition the objects by dragging the corresponding dots to adjust the [x, y] coordinates as needed.

\begin{figure}
    \centering
    \includegraphics[width=0.95\linewidth]{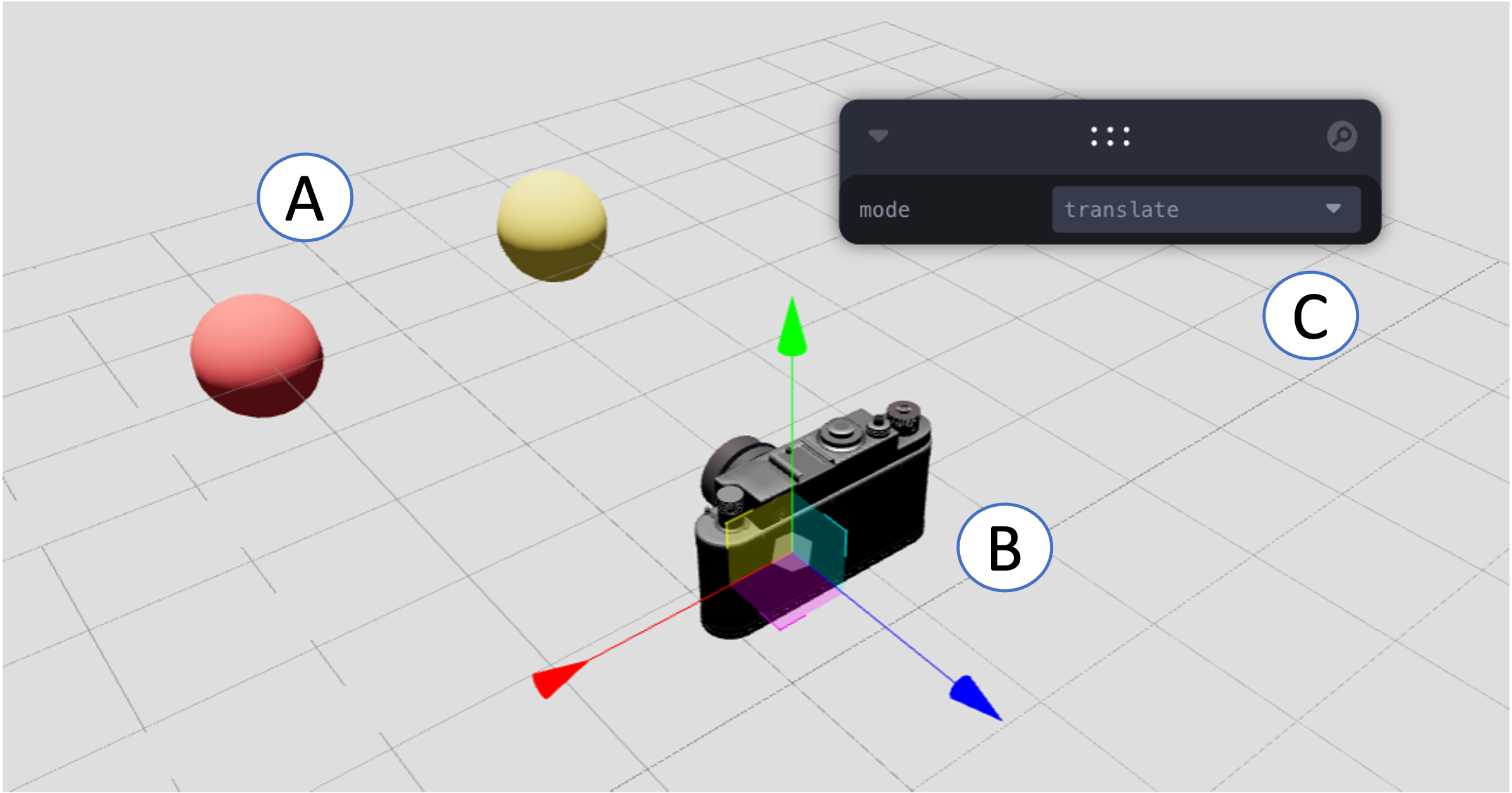}
    \caption{(A) 3D sound source manipulation, where users can adjust the spatial position of each sounding object by moving the corresponding sphere. (B) Users can change the viewing point position using the camera object. (C) Users can choose to move or rotate objects, especially the camera object.}
    \label{fig:teaser-C}
    \Description[]{The figure illustrates the function of each widget in the system interface. (A) 3D sound source manipulation. Users can alter the spatial position of each sounding object by moving the corresponding sphere. (B) Users can manipulate the position of the viewing point through the camera object. (C) Users can specify if they want to move or rotate the object (especially for the camera object).}
\end{figure}

In contrast, the 3D manipulation panel (Fig.\ref{fig:teaser-C}) operates within a simulated 3D space, independent of the original video frame. Predicted 3D positions of sounding objects are represented as colored spheres. The panel gives users a greater degree of creative freedom when manipulating the positions of sounding objects in several ways. Users can manipulate objects within the open 3D space, modify the point of recording (Fig.\ref{fig:teaser-C}-B) through rotation or translation (Fig.~\ref{fig:teaser-C}-C), and rotate the entire simulated space to view object positions and the camera from various angles.

The goal of the 2D and 3D direct manipulation panels is to allow users selecting preferred reference systems for balancing the trade-off between error correction and creative support. Specifically, the 2D panel allows users to easily correct errors in predicted spatial positions by aligning the dot with the actual sounding object in the video frame.  However, it limits depth correction and experimentation with spatial audio effects from different perspectives, thus restricting creativity. The 3D panel, on the other hand, replaces the original video frame reference with a more flexible 3D coordinate system and viewing point, allowing users to manipulate objects within an open space and incorporating diverse interactive strategies to support creativity.

\subsubsection{Video panel}
The video panel displays the video and includes several interactive elements for user control.

First, users can select the audio channel mode from a drop-down list, as seen in Fig.~\ref{fig: teaser-A}-B. Mimosa supports four modes: monaural (1-channel), stereo (2-channel), quadraphonic (4-channel), and 5.1-channel audio.

The number of channels corresponds to the number of volume indicators in Fig.~\ref{fig: teaser-A}-A. Each bar represents the sound intensity of its respective channel. For example, in a 5.1-channel setup, the bars at the upper right, center, upper left, bottom left, and bottom right indicate the sound intensity for the front left, center, front right, rear left, and rear right channels, respectively. The subwoofer's sound intensity is not displayed, as it does not convey directional information. These channel volume indicators aid users in the error discovery process by visually representing sound intensity across each channel.

Additionally, the toggle in Fig.~\ref{fig: teaser-A}-C allows users to enable or disable the use of model-predicted spatial coordinates. When enabled, the system automatically interpolates the position of a sounding object between two frames where the user modifies the positions, reducing the effort required for manual editing of individual frames. When disabled, the spatial coordinates of the sound source remain unchanged until the user moves the object, providing full user control.

Lastly, users can control video playback using the control buttons in this panel, as displayed in Fig.~\ref{fig: teaser-A}-D.

\subsubsection{Audio properties display and control panel}
\label{sec:audio_property}
\begin{figure}[!htb]
    \centering
    \includegraphics[width=\linewidth]{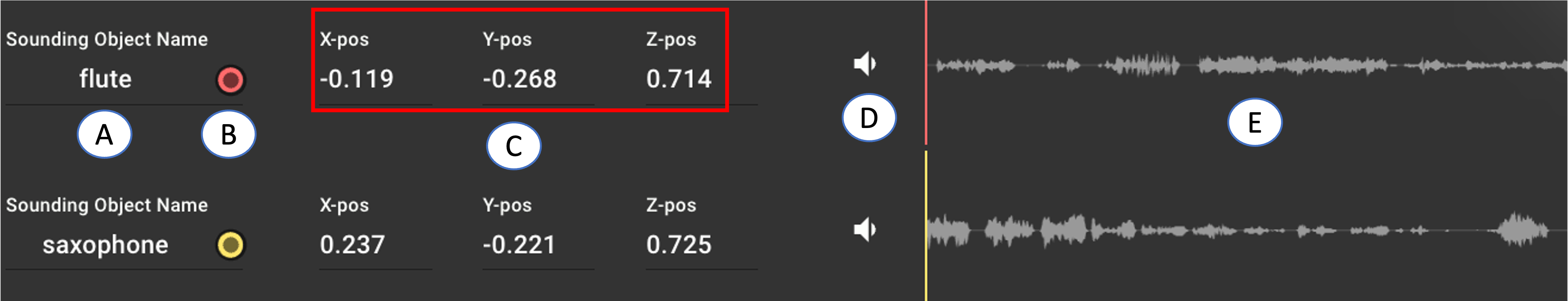}
    \caption{(A) Users can specify the audio-visual correspondence by changing the name of the sounding object. (B) Object color indicator allows users to change the color of dots (in 2D manipulation) or spheres (in 3D manipulation) representing sounding objects. (C) Numeric spatial position enables users to view and modify the numeric coordinates of sounding objects. (D) By clicking the icon, users can control the volume through a volume slider. (E) Sound waveform display that visualize the current audio.}
    \label{fig:teaser-B}
    \Description[]{The figure illustrates the function of each widget in the system interface. (A) Users can specify the audio-visual correspondence by changing the name of the sounding object. (B) Object color indicator. Users can change the color of the dots (in 2D manipulation) or spheres (in 3D manipulation) that represent the sounding objects. (C) Numeric spatial position. Users can view and change the numeric coordinates of the sounding objects. (D) By clicking the icon, users can control the volume through a volume slider. (E) Sound waveform display.}
\end{figure}

This panel allows users to correct misalignments between separated soundtracks and corresponding visual objects. By default, the identified sounding object name appears in Fig.\ref{fig:teaser-B}-A. Users can select the correct visual object name from a drop-down menu by clicking on the name field if the default name is incorrect. Additionally, users can click the \faIcon{bullseye} button (Fig.\ref{fig:teaser-B}-B) to change the color of each sounding object's visual representation in the 2D and 3D manipulation panels.

The panel also displays the spatial location of individual sounding objects in a numeric format (Fig.~\ref{fig:teaser-B}-C). Users can modify the position of each object by directly editing the corresponding numeric value.

Moreover, users can refer to the waveform of each separated sound source (Fig.\ref{fig:teaser-B}-E) to facilitate navigation to specific timestamps in the video. The volume of each sound source can be adjusted by clicking the \faIcon{volume-down} button (Fig.\ref{fig:teaser-B}-D), which activates a volume slider.

\subsection{Pipeline for Audio-Visual Spatialization}
\label{sec:voasFramework}

In this section, we introduce our audiovisual pipeline (Fig.~\ref{fig:ml_pipeline}) for generating spatial audio effects for each sounding object.

\subsubsection{Inferring the 3D positions of the visual objects in 2D videos}
\label{sec:3dcoordinate}
To achieve this, the video is first sampled into individual frames. A pre-trained object detector is then used to extract the 2D positions and respective categories of objects within each video frame (Fig.\ref{fig:ml_pipeline}-B). We adopt a pre-trained image depth estimation model\cite{kim2022global} to generate a pixel-level depth map for each video frame (Fig.~\ref{fig:ml_pipeline}-A). Using the object's 2D position from the previous step, we query the depth map to obtain the depth information for each object. This approach allows us to accurately determine and track the 3D position of each object over time.

The object detector used in \textsc{Mimosa} was built using the Faster R-CNN neural network architecture~\cite{ren2015faster} and implemented with the Detectron2 framework\footnote{\url{https://github.com/facebookresearch/detectron2}}. The model was pre-trained on the MS-COCO dataset~\cite{lin2014microsoft}, a widely used benchmark dataset for object detection in computer vision. The object detector does not require fine-tuning when integrated into our system.

\subsubsection{Separating individual soundtracks and aligning them with visual counterparts}
At the audio level, a sound separation model~\cite{tian2021cyclic,wisdom2021s} is used to extract individual soundtracks from the original sound mixture. Each soundtrack contains sound from a single sounding object (Fig.~\ref{fig:ml_pipeline}-C).

To align auditory information with visual information obtained in Section~\ref{sec:3dcoordinate}, an audio tagging model~\cite{kong2020panns} (Fig.~\ref{fig:ml_pipeline}-D) is used to predict the sound types for each sound source (e.g., speech, telephone, music, etc.). The separated soundtracks are then mapped to the corresponding category of visual objects, using their names as bridges.

\subsubsection{Real-time spatial audio rendering}
\label{sec:audio-rendering}
This module takes the separated soundtrack for each sound source, mixes them together based on the corresponding 3D positions of each object, and distributes the mixed sound signal into different channels depending on the speaker setup.

The module is built using the \texttt{PannerNode} object in the WebAudio API\footnote{\url{https://developer.mozilla.org/en-US/docs/Web/API/Web\_Audio\_API}}, which simulates sound effects in a simulated 3D space given the 3D coordinates of the sounding object and the listening position. Since the original \texttt{PannerNode} only supports two-channel audio rendering, we combined two \texttt{PannerNodes} to compute spatial effects separately for the sounding sources from the front and the back.

\subsection{Implementation}
\label{sec:implementation}
The front-end of \textsc{Mimosa} is hosted on a virtual machine through Google Cloud. The Adobe Premiere Pro plugin for \textsc{Mimosa} is built with the Adobe Common Extensibility Platform (CEP)\footnote{\url{https://github.com/Adobe-CEP}}. \textsc{Mimosa}'s front-end UI was implemented in React\footnote{\url{https://reactjs.org}} rather than using the native CEP APIs in Premiere Pro to enable flexible interactions. The backend spatializing pipeline runs on a workstation with an AMD Ryzen Threadripper 3960X CPU and an NVIDIA RTX A6000 GPU. 

\section{Technical evaluation}
In this independent evaluation, we want to answer two questions: 
\begin{itemize}
    \item[(i)] How effective is the audiovisual sound spatialization pipeline against the offline end-to-end model?
    \item[(ii)] How is the quality of the spatial audio effect generated by users using \textsc{Mimosa}?
\end{itemize}

\subsection{Dataset}
We recorded six sample videos across various scenarios. The videos were recorded with original spatial audio effects, serving as one of the contrast conditions in subsequent analyses. Each video is approximately one minute long to maintain a balance between the overall duration of the user study and the number of videos participants we can test. Detailed information on each video is shown in Table~\ref{tab:scenarios}.

Each video clip with the ground-truth spatial audio was recorded with an iPhone 13 camera and a ZOOM H3-VR microphone\footnote{\url{https://zoomcorp.com/en/us/handheld-recorders/handheld-recorders/h3-vr-360-audio-recorder/}}, a popular economic 360\degree~audio recorder with a four-capsule ambisonic microphone array. \looseness=-1
\begin{table}
\centering
\begin{tabular}{c|c|c|c} 
\toprule
\multirow{2}{*}{\textbf{Video ID}} & \multirow{2}{*}{\textbf{Video Scenario}} & \multirow{2}{*}{\textbf{Duration}} & \textbf{Number of} \\
& & & \textbf{Sound Sources}\\ 
\hline
V1   & Vehicle honking         & 0:52              & 2                             \\
V2   & Man speaking            & 0:54              & 2                             \\
V3   & Music duet              & 0:43              & 2                             \\
V4   & Man playing basketball  & 1:26              & 2                             \\
V5   & Dog barking             & 1:06              & 2                             \\
V6   & Cooking in the kitchen   & 1:37              & 4                             \\
\bottomrule
\end{tabular}
\caption{Summary of the video details for the user study.}
\label{tab:scenarios}
\end{table}

\subsection{Evaluation Method}
We recruited eight independent evaluators to subjectively evaluate different types of audio. An evaluator viewed a video with an associated audio type at a time. The evaluators were 20 to 30 years old and were recruited from the local community. They had no previous experience with \textsc{Mimosa} and did not participate in the user study, ensuring that their ratings remained unbiased. Each evaluator reported having experience consuming videos with and without spatial audio effects.

Each video has five different audio types: 

\begin{itemize}
    \item[(1)] Raw audio (RA): the pre-recorded original spatial audio.
    \item[(2)] Monaural audio (MA): The pre-recorded original audio played in a single-channel setting (no spatial effect).
    \item[(3)] Spatial audio generated by an offline end-to-end model~\cite{xu2021visually} (OA). \looseness=-1
    \item[(4)] Default-generated spatial audio by \textsc{Mimosa}'s pipeline without any user intervention (DA).
    \item[(5)] Spatial audio created by users using the \textsc{Mimosa} system (UA).
\end{itemize}

We shuffled the order of video clips created by participants (UA) along with the other four versions (RA, MA, OA and DA) and presented them to the evaluators in random order. Each evaluator rated about 12 videos, each with five different audio types. 

Evaluators were asked to rate each video on \texttt{Immersion} and \texttt{Realism} separately on a 7-point semantic differential scale. \texttt{Immersion} evaluates the extent to which the spatial audio effect was perceptible, and how well the spatial locations of objects could be inferred from the audio effect. \texttt{Realism} focuses on identifying any distortion or dissonance that could be perceived as unreal. 

\begin{table*}
\centering
\begin{tblr}{
  cells = {c},
  cell{1}{3} = {c=5}{},
  cell{3}{1} = {r=7}{},
  cell{10}{1} = {r=7}{},
  hline{1,3,17} = {-}{0.08em},
  hline{2} = {3-7}{},
  hline{10} = {-}{},
  rowsep = 0.1ex,
}
                    &                   & Audio type             &                   &                             &                               &                              \\
\textbf{Metric}     & \textbf{Video ID} & \textbf{Monaural (MA)} & \textbf{Raw (RA)} & \textbf{Offline model (OA)} & \textbf{\textsc{Mimosa} default (DA)} & \textbf{User generated (UA)} \\
\textbf{Immersion } & V1                & 2.56                   & 4.5               & 3.25                        & 4.81                          & 6.12                         \\
                    & V2                & 1.69                   & 5.69              & 2.36                        & 5.12                          & 6.31                         \\
                    & V3                & 1.5                    & 4.88              & 4.69                        & 4.25                          & 6.00                         \\
                    & V4                & 2.06                   & 3.81              & 2.25                        & 5.00                          & 6.25                         \\
                    & V5                & 1.63                   & 4.69              & 2.06                        & 4.38                          & 5.50                         \\
                    & V6                & 2.25                   & 3.5               & 1.94                        & 3.12                          & 5.94                         \\
                    & Average           & 1.95                   & 4.51              & 2.76                        & 4.47                          & 6.03                         \\
\textbf{Realism }   & V1                & 5.94                   & 5.69              & 3.81                        & 3.06                          & 5.63                         \\
                    & V2                & 5.69                   & 6.31              & 3.44                        & 3.13                          & 5.63                         \\
                    & V3                & 5.44                   & 5.81              & 4.81                        & 3.63                          & 5.38                         \\
                    & V4                & 5.75                   & 6.06              & 4.38                        & 4.19                          & 5.44                         \\
                    & V5                & 5.38                   & 5.88              & 3.25                        & 3.75                          & 5.88                         \\
                    & V6                & 5.88                   & 6.44              & 4.06                        & 4.25                          & 5.56                         \\
                    & Average           & 5.68                   & 6.03              & 3.96                        & 3.67                          & 5.58                         
\end{tblr}
\caption{Average rating scores of immersion and realism for videos with varied audio types by external evaluators.}
\label{tab:immersion_realism}
\end{table*}

\subsection{Results}
The quantitative rating results are shown in Table~\ref{tab:immersion_realism}. The value of each cell represents the average rating for the corresponding audio type in the video. We found \textbf{videos with spatial audio effect after user editing with \textsc{Mimosa} (UA) were more immersive than other types of audio effects.} We adopted the Friedman test to measure the difference among the five audio types. The result showed that there was a significant difference among the mean immersion scores of various audio types  ($p<0.001$). Furthermore, we compared the difference between every two audio types with Wilcoxon signed-rank test. The result shows that the differences between the mean of UA and all other audio types are also statistically significant ($p<0.001$).

Additionally, we found that immersion score of DA ($\mu=4.47$; $\sigma=0.94$) was close to RA ($\mu=4.51$; $\sigma=1.19$). The difference of ratings between DA and RA is not significant ($p=0.54$). The results suggest that spatial audio effects generated using \textsc{Mimosa}'s pipeline, even in a fully automated way, achieve an immersive experience comparable to that of raw audio recorded with a 360-degree audio recorder. The user's validation and revision in \textsc{Mimosa} further improved the immersion of the generated spatial audio effects. Noticeably, the default-generated effects are better in \textit{realism} and \textit{immersion} than OA, which is generated by an end-to-end ML model.\looseness=-1

Last, we found \textbf{spatial effects after user editing with \textsc{Mimosa} (UA) compromised realism compared to raw audio (RA) but show improvement over the fully-automated outcome (DA).} We identified a significant difference among various audio types through a Friedman test in \texttt{Realism} ($p<0.001$). Specifically, RA received the highest mean score ($\mu=6.03$; $\sigma=0.59$), followed by MA ($\mu=5.68$; $\sigma=0.33$). This observation suggests that the computationally generated spatial audio inevitably introduces distortion, leading to a decrease in user-perceived realism.

\section{User Study}
Informed by prior research in measuring creativity support and heuristic evaluation of digital tools~\cite{cherry2014quantifying,nielsen1990heuristic}, we conducted an in-person lab study\footnote{The study protocol has been reviewed and approved by the IRB at our institution.} to evaluate the usability, usefulness, and user experience of \textsc{Mimosa}. The study seeks to answer the following research questions.

\begin{itemize}
    \item[RQ1:] How usable is \textsc{Mimosa} in assisting users to edit spatial effects and support their expressiveness?
    \item[RQ2:] What is the user perception towards the interactive strategies of \textsc{Mimosa}?
    \item[RQ3:] How would users utilize \textsc{Mimosa} to augment the spatial audio effects while editing?
\end{itemize}

\subsection{Participants}
15 participants (9 men, 6 women) aged 20--30 were recruited from the local community to use the \textsc{Mimosa} system to generate spatial audio effects for videos. 8 participants were amateur video creators who had posted their videos on websites such as YouTube, while the other 7 did not consider themselves video content creators. 6 participants were experienced with video editing tools such as Adobe Premier Pro; 7 were novice users of these tools, and the rest 2 had no video editing experience. Each participant was compensated with \$15 USD for their time.

\subsection{Study Procedure}

\begin{figure*}[t]
\centering
  \includegraphics[width=0.85\linewidth]{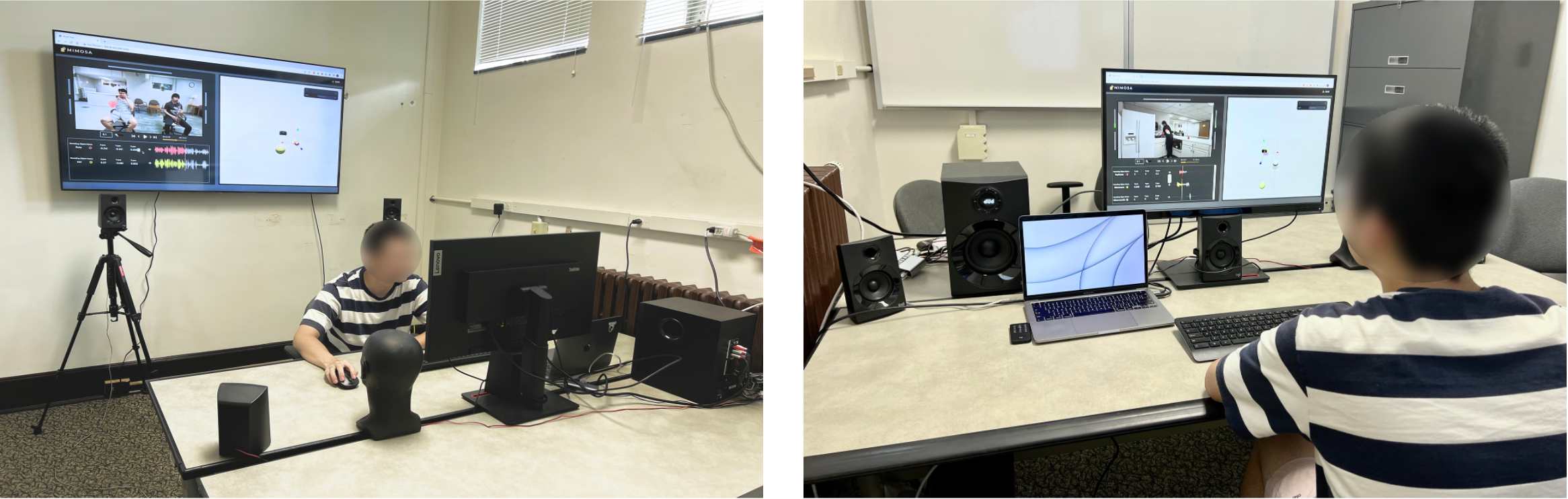}
  \caption{The setup of the user study.}
  \label{fig:study}
  \Description[]{This picture shows the setup of the user study. The participant sits in the middle of a 5.1-channel speaker, using MIMOSA through a laptop with a 24-inch monitor.}
\end{figure*}

The study was conducted in a usability lab (Fig.~\ref{fig:study}). The lab is equipped with a set of Logitech Z606 5.1 surround speakers\footnote{\url{https://www.logitech.com/en-us/products/speakers/z606-surround-sound-system.980-001328.html}} that provided audio playback during the study.

At the beginning of each session, the participant signed the consent form and completed a demographic questionnaire. After the researcher gave a brief introduction to the study, the participant watched a five-minute tutorial video on how to use \textsc{Mimosa} and a one-minute 5.1 channel spatial audio test video to check the surround speakers. 

In the study session, each participant completed two tasks for each of the six video clips. The order of the six video clips was randomized, but the tasks for each video clip followed the same order.  \looseness=-1

The first task seeks to evaluate \textsc{Mimosa}'s capability in facilitating effective human-AI collaboration for users to validate model-generated results and repair any errors they find. In the first task, the user was asked to create \textit{realistic} spatial sound effects for the video by repairing the inaccuracy and errors in the video with default generated spatial audio (DA) using \textsc{Mimosa}. 

The second task evaluates \textsc{Mimosa}'s capability to support flexible customization of spatial audio effects. To familiarize the participant with creation process, we demonstrated the function by guiding the users to create two simple scenarios. The two scenarios were: \textit{ ``Please move the Saxophone player so that it gradually moves away from you.''}, \textit{``Please turn your face back to the basketball player.''} The researcher would answer any questions from the participants during this warm-up process. After creating the effects according to the provided instructions, the user was asked to use \textsc{Mimosa} to create a customized spatial audio effect to their own liking for the current video clip.

\begin{figure*}
  \centering
  \includegraphics[width=0.9\linewidth]{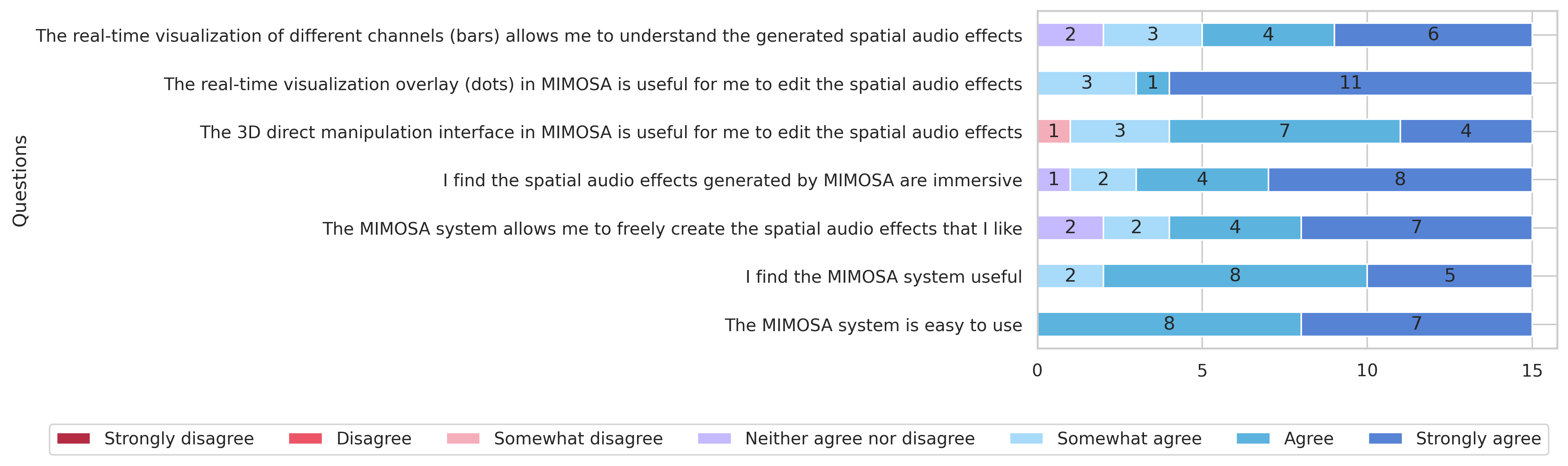}
  \caption{Results of the post-study questionnaire}
  \label{fig:questionnaire}
  \Description[]{Results of the post-study questionnaire. The diagram shows that most participants found MIMOSA useful in generating immersive spatial audio and easy to use.}
\end{figure*}

The study session ended with a post-study questionnaire and a 10-minute semi-structured interview. In the questionnaire, participants rate statements on the usability and user experience of \textsc{Mimosa} and its key interaction features on a seven-point Likert scale from \textit{``strongly disagree''} to \textit{``strongly agree''} (The questions and results are shown in Fig.~\ref{fig:questionnaire}). During the interview, we asked follow-up questions about their responses to the post-study questionnaire, especially when they gave negative ratings to any aspects of \textsc{Mimosa}. Following established open coding methods~\cite{braun2006using,lazar2017research}, an author conducted a thematic analysis of interview transcripts to identify insights and findings particularly related to user experiences, challenges, system usability, and suggestions for new features using an inductive approach.

All 15 participants successfully completed both tasks for all six video clips. All user study sessions were video recorded with the consent of the participants.

\subsection{Results and Findings}
\label{sec:qualitative_results}

Overall, participants were satisfied with their experience with \textsc{Mimosa}. They found it easy to use, useful, expressive, and capable of generating immersive spatial audio effects. Specifically, \textsc{Mimosa} scored 6.47 ($\sigma=0.52$) in \textit{``\textsc{Mimosa} is useful''}, 6.20 ($\sigma=0.68$) in \textit{``the spatial audio effects created through \textsc{Mimosa} is immersive''}, 6.27 ($\sigma=0.96$) in \textit{``\textsc{Mimosa} allows me to freely create the spatial audio effects that I like''}, and 5.87 ($\sigma=1.06$) in \textit{``\textsc{Mimosa} is easy to use''}. Regarding the effectiveness of \textsc{Mimosa}'s interaction features, the 3D direct manipulation panel scores 6.07 ($\sigma=1.09$), and the 2D dots visualization overlay scores 6.53 ($\sigma=0.83$).

\vspace{0.5em}
\noindent\textbf{Generate spatial audio effects with ease. }Participants stated that they could \textit{``quickly get familiar with the functions of the system.''} (P1) and \textit{``speed up the editing process after finishing editing the first several videos.''} (P13). Additionally, most participants mentioned that the spatial effects after editing were more immersive. When asked about the factors contributing to the immersion, participants provided examples such as  \textit{``I can clearly feel the car is moving from left to right.''} (P8); and \textit{"When I moved the Saxophone to my back, the sound was actually coming from that position."} (P3). 

However, several participants noted issues, such as the back-to-front sound transition not being as natural as expected. Users reported hearing a \textit{``crunch sound''} (P9) during playback. Additionally, some users mentioned that the visual position of the sounding objects in the interface did not always align with the auditory position they perceived in the playback. For example, P13 mentioned that \textit{``In one of the videos, I wish I could move the sound source further away, but even though I did it in the interface, the sound I heard was still louder than I expected.''}

\vspace{0.5em}
\noindent\textbf{Support creativity and expressiveness by intuitive manipulation and real-time playback. }Informed by the existing evaluation dimensions for creativity support tools~\cite{cherry2014quantifying}, we asked users about how enjoy, expressive, and immersive when they were using the tool, and how helpful the tool was in allowing them tracking different ideas. Participants expressed that they could easily \textit{``test a variety of different audio settings''} (P14) due to the flexible manipulation strategies offered by \textsc{Mimosa}. P11 also mentioned that \textit{``Aligning the dots and the sounding objects in the video frame cost nearly no labor to me, so I felt really excited playing around with different settings.''} Participants also appreciated the flexibility of augmenting existing spatial effect by intentionally \textit{``Increase the distance between the two sounding objects or bring them closer to the reference point.''} (P3). For the 3D manipulation panel especially, participants applauded it as it allows for flexibly changing the viewing point, which offered them more creativity control. For example, P13 stated that \textit{``The 3D panel allows me to listen to different spatial effects from varied perspectives in a simulated 3D space by moving the viewing point.''} More specifically, P15 stated that \textit{``It (3D/manipulation panel) allows me to imagine I can walk into the video scene and pretend to be at a particular position to hear the music instruments playing.''} Additionally, a number of participants emphasized that being able to hear the effects after they made changes further encourages them to test more scenarios due to the minimized cognitive load. For instance, P8 expressed that: \textit{``getting instant feedback is super important in the editing process...this allows me to do the editing and evaluating tasks simultaneously.''}

\vspace{0.5em}
\noindent\textbf{Handle audio error by using visual clues. }We found participants who were amateur content creators relied on using the visual clues to discover and repair errors in the generated spatial audio effects. We observed that most users first discovered the inaccuracy in the generated spatial effects by noticing the difference between the position of the colored dots and the corresponding visual objects. For example, P13 noted that comparing with listening to the generated spatial audio and trying to identify the errors, by looking the misalignment of the representative dots overlay, he could \textit{``easily found out unnoticed errors by just `hearing'.''} P15 also mentioned that:\textit{``(the strategy) makes fixing the error easy and intuitive, as I just need to drag the dots to make it overlap with the objects.''} The user feedback suggests that \textsc{Mimosa} is able to augment the audio perception and understanding of the amateur creators when perceiving the spatial audio effect, achieved by providing the associated visual information.

\vspace{0.5em}
\noindent\textbf{Selectively choose the manipulation method depending on personalized user needs. } \textsc{Mimosa} provides three distinct approaches for participants to manipulate the location of sounding objects in space (Figure~\ref{fig: teaser-A}-E, Figure~\ref{fig:teaser-C}, and Fig.~\ref{fig:teaser-B}-C). Most participants found manipulating sounding objects using 2D or 3D manipulation panels by \textit{dragging} was more user-friendly and effective than editing the numerical values of their coordinates. For instance, participant P2 noted that:\textit{``I only used the numerical values as a reference, because the spatial effect changes were more immediate when directly moving the dot or spheres.''} However, some participants preferred numerical input as the primary editing method. Participant P1, for example, stated that: \textit{``I felt more confident editing spatial audio effects when using the numerical input, especially when I tried to create a new spatial audio effect and wanted to position different objects at various places.''} The result suggests that users have various preferences in choosing the way of manipulation. The decision depends on whether they prefer a more deterministic way or a free-form one. Noticeably, there are some participants (P6, P10, P11) mentioned that they do not have significant preferences but to mix-use all the functions.

\vspace{0.5em}
\noindent\textbf{Suggestions for new features. }During the interviews, participants offered feedback and recommendations for potential new features in \textsc{Mimosa}. P8 proposed the addition of a separate drop-down menu within the system, allowing users to select specific objects they wish to manipulate. This menu would facilitate distinguishing between various audible objects when their corresponding spheres (in the 3D panel) or dots (in the 2D panel) overlap. Another concern expressed by participants was the difficulty in accurately navigating back to previous actions. P7 suggested incorporating a new panel that records actions along with their corresponding timestamps. This addition would enable users to effortlessly review and modify their action history, consequently enhancing the editing efficiency.
\section{Discussion}
\subsection{Error Discovery and Repair as a Foundation for Human-AI Co-Creation}
\label{sec:disErr}
Effective human-AI collaboration requires a clear understanding of mutual goals ~\cite{wang2020human} and trust between the human and the AI system~\cite{rebensky2021whoops}. However, most machine learning models will inevitably introduce errors into predictions that threaten mutual understanding and trust, making human-AI collaboration systems uniquely difficult to design~\cite{10.1145/3313831.3376301}. The study results of \textsc{Mimosa} show that an effective error discovery and repair mechanism is crucial in the context of human-AI co-creation where each party needs to work closely with the other party and build upon each other's intermediate results. 

The design of \textsc{Mimosa} exemplifies the use of error discovery and repair as a foundation to support effective human-AI co-creation. Instead of training an end-to-end model to predict spatial audio, \textsc{Mimosa} uses a multi-step pipeline that produces useful and interpretable intermediate results at each of its steps for users to review, validate, and edit the model results. For example, the 2D panel overlays the results of model-inferred sounding objects on top of the original video, allowing users to easily validate whether the inferred positions align well with the underlying visual objects and repair them through direct manipulation.  As discussed in Section~\ref{sec:qualitative_results}, many study participants found that the interaction strategies in \textsc{Mimosa} made it easy for them to discover and repair the errors. The effectiveness of such a method was also evidenced in the improvement in \textit{realism} from DV to EV in our evaluation.

The success of \textsc{Mimosa} in error discovery and handling demonstrates the adoption of classic theories in multi-modal interactions~\cite{10.1145/319382.319398,10.1145/302979.303163} in the new context of human-AI co-creation of multimedia contents. Specifically, the design of \textsc{Mimosa} utilizes the \textit{mutual disambiguation} paradigm. Because the generated result of the model is auditory, it is more difficult and cognitively demanding for the user to validate whether it aligns well with the visual information in the original video. To address this problem, \textsc{Mimosa} simultaneously represents the result of generated spatial audio effects visually on the 2D/3D panels, allowing users to easily compare its alignment with the visual information in the video. To make edits, the user can either manipulate it visually (i.e., moving the visual indicator) or adjust its auditory properties directly through the soundtrack information panel. The availability of both interaction modalities at the same time enables the user to choose whichever modality is most natural for their task goals (as we learned from observations and interviews from the user study), while the synchronous representation of the result in both modalities allows the users to easily identify issues in either modality. \looseness=-1

\subsection{Going Beyond the ``Ground Truth''}

Compared to traditional media editing tools, creative tools often promise users full expressiveness that allows them to flexibly explore the creative space beyond ``realism''. However, this characteristic poses significant challenges when incorporating machine learning models in creative tools for human-AI co-creation: These models were built to predict the results of ``ground truth'' learned from the training data. Although some generative models such as~\cite{louie2020novice,reed2016generative} and recent commercial systems such as DALL-E\footnote{\url{https://openai.com/dall-e-3/}} can generate impressive ``artistic'' work on their own, they lack the support of user control, which is necessary for enabling true co-creation.\looseness=-1

The design of \textsc{Mimosa} illustrates an approach to support user-initiated augmentation and creation based on model predictions. Throughout the pipeline, the model's strategies to generate and present its output are aligned with the user's existing mental model and workflow of the task. For example, the 3D panel visualizes the positions of each inferred sounding object and the reference of viewing point (camera) in a simulated 3D space. The user can either move the sounding objects, move the camera, or point the camera in a different direction (changing the viewing angle). All these operations provide close analogies to the manipulation of real-world objects and therefore pose a lower learning curve to the users. The user can also easily amplify the presence of a particular sounding object by directly manipulating the intensity of its sound in the sound track information panel. As a result, users can easily understand and leverage the model predictions in this specific context. In this way, users can offload laborious lower-level sub-tasks to ML models and focus on pursuing higher-level creation goals.

\subsection{Human-AI Collaboration for Augmentation of Multimedia Content}
\textsc{Mimosa}'s approach to creating spatial audio for videos with only mono or stereo audio illustrates a approach that computationally \textit{augments} existing multimedia contents to a new format with richer information through human-AI co-creation. The AI model can predict the outcome of the new format, while users can help validate and fix the alignment between the old format and the new format. 

We plan to continue exploring this approach in other multimedia domains such as augmenting regular 2D video content into AR/VR content in 3D. Content creation for VR/AR faces similar challenges to what we encountered in our problem domain---it requires specialized recording equipment and significant creator expertise. While there are millions of existing 2D video footage available, they cannot be easily used in the 3D VR/AR space. Although ML models are available for synthesizing 3D contents from 2D ones~\cite{mildenhall2021nerf,xie2016deep3d,konrad2013learning}, the limited accuracy and applicability hinder their practical use. Besides, another content creation domain that can benefit from this approach is authoring audio descriptions for video content. While AI models like~\cite{10.1145/3411764.3445347} can generate scene descriptions for video content in an end-to-end fashion, they fall short on accuracy, coherence, naturalness, and context awareness~\cite{10.1145/3441852.3476550}, which could be fixed by human validation and repair through a human-AI collaboration workflow~\cite{10.1145/3357236.3395433}. An easy-to-use and effective human-AI collaboration framework like \textsc{Mimosa} would also make AI assistance in these domains of content creation more accessible to amateur creators. \looseness=-1

\section{Limitations \& Future work}
The current version of \textsc{Mimosa} presents several technical and user study limitations for future work.

\vspace{0.5em}
\noindent\textbf{Support for more general video types.} Due to the limited support for different topic domains in its sound separation, audio tagging, and object detection models, \textsc{Mimosa} currently supports only a limited range of sounding objects. Additionally, if a specific type of sounding object is not present in the training stage of the model, the model would struggle to recognize the visual object, separate its soundtrack, and predict its sound type. A promising approach to address this limitation is to incorporate an active learning approach that allows ML models to incrementally learn from the user's manipulation in real-time, thereby improving the model's performance. 

\vspace{0.5em}
\noindent\textbf{Model the interaction between the sound and environment. }Currently, the spatialization strategy used by \textsc{Mimosa} only considers the spatial position of the sounding object, without accounting for other factors such as reflection, absorption, and diffraction of sound. 
While new algorithms have been proposed for this problem~\cite{raghuvanshi2014parametric,raghuvanshi2018parametric}, we expect it to remain a challenge in the near future due to the level of complexity involved. These algorithms require extensive information about the scene's geometry and material properties, which may not be readily available or easily inferred from videos. 

\vspace{0.5em} 
\noindent\textbf{Handle out-of-sight objects. }The current version of \textsc{Mimosa} is limited to handling sounding objects that are \textit{visible} in the video frame. When the object detector is out of sight, the spatialization pipeline fails to work. To address this issue, \textsc{Mimosa} currently employs linear interpolation for transient out-of-sight situations within video clips. In other cases, the system relies on users to manually specify the locations of sounding objects, which can be time-consuming and labor-intensive. While it is challenging to infer the location of objects that are never in sight, in future versions of \textsc{Mimosa}, we will explore the use of state-of-the-art computer vision techniques in trajectory prediction~\cite{wang2022stepwise,mohamed2022social,scibior2021imagining,liang2020simaug} to predict in-between positions for objects that are \textit{temporarily} out-of-sight due to their movements. \looseness-1

\vspace{0.5em}
\noindent\textbf{Deployment study} As part of our future work, we plan to conduct a field deployment study involving content creators using \textsc{Mimosa} in their own video projects. Due to the constrained duration of the lab study, we did not systematically assess how users could integrate the tool into their actual creative workflows, such as incorporating Premiere Pro (Pr), although \textsc{Mimosa} supports an integrated workflow as a Pr plug-in. A deployment study will enable us to examine the long-term in-situ usage of \textsc{Mimosa} within its intended context. This will also allow us to assess the real-world effectiveness of the system for content creators and verify its ecological validity.

\section{Conclusion}
In this paper, we presented \textsc{Mimosa}, a human-AI co-creation tool that enabled amateur users to computationally generate and interactively manipulate spatial audio effects in videos that only had monaural or stereo audio. \textsc{Mimosa} featured a human-AI collaborative spatializing pipeline that produces user-interpretable and controllable intermediate results to support effective error discovery and repair. A controlled user study of \textsc{Mimosa} demonstrated that \textsc{Mimosa}'s approach could support amateur content creators to generate immersive and realistic spatial audio and enable the flexible creation of customized spatial effects. Our findings provided design implications for AI-assisted content creation and future human-AI collaboration tools for working with multi-modal data.

\begin{acks}
This work was supported in part by an AnalytiXIN Faculty Fellowship, an NVIDIA Academic Hardware Grant, a Google Cloud Research Credit Award, a Google Research Scholar Award, and the NSF Grant 2211428.
\end{acks}

\balance
\bibliographystyle{ACM-Reference-Format}
\normalem
\bibliography{mimosa_references}

\end{document}